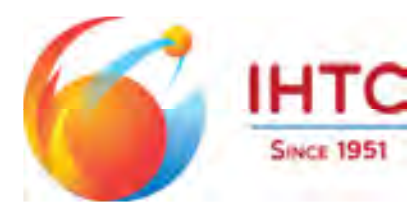

# A MULTISCALE STUDY OF FILM THICKNESS DEPENDENT FEMTOSECOND LASER SPALLATION AND ABLATION

**Pengfei Ji[1,2,3], Yuwen Zhang[2*], Yiming Rong[1], Yong Tang[3]**
[1]Southern University of Science and Technology, Shenzhen 518055, China
[2]University of Missouri, Columbia, MO 65211, USA
[3]South China University of Technology, Guangzhou 510640, China

## ABSTRACT

A multiscale studying integrating *ab initio* quantum mechanics, classical molecular dynamics and two-temperature model, is carried out to study film thickness dependent femtosecond laser spallation and ablation. As an interval of $130.73\,nm$, five silver films with increasing thickness from $392.19\,nm$ to $915.11\,nm$ are simulated. Absorbed laser fluences of $0.1\,J/cm^2$ and $0.3\,J/cm^2$ are chosen to observe the laser spallation and ablation. The simulation results show that film thickness has close correlation with the Kelvin degree of heating of the laser irradiated silver films, which further affects femtosecond laser spallation and ablation. Suggestions for precise micromachining are proposed in this paper.

## 1. INTRODUCTION

Femtosecond laser material process is a widely acknowledged approach in micromachining and microfabrication [1–3]. Properly setting up the parameters of laser pulse and studying the optical, mechanical and thermophysical properties of material, help to achieve smoothly manufactured surface and highly precise control in femtosecond laser processing of thin silver film. Considerable attentions have been drawn in studying the ultrafast laser interaction with metal film. The rapid melting and resolidification of gold film was simulated by combining the two-temperature model with the interfacial tracking method [4]. The effects of laser pulse width and fluence were studied [5]. By imposing the single pulse, multiple pulse and pulse train, Huang et al. carried out numerical simulations to investigate the effects of film thickness on laser melting and vaporization [6]. The one-dimensional and two-dimensional gold thin film gratings were fabricated with the help of nanosecond laser induced thermos-elastic force to detach the film from substrate [7].

There were a bunch of mechanisms interpreting the ultrafast laser spallation and ablation. The thermal mechanism stated that the high rate of vaporization of thermalized material and formation of plasma induced the removal of laser irradiated material [8–10]. The mechanical mechanism explained that the removal of material resulted from the generation of tensile stress [11,12]. The Coulomb explosion mechanism couples the laser excited electronic energy from intense electromagnetic fields into atomic motion, which leads to the metal film explodes into plasma of ionized atoms [13–15]. The hot-electron blast force mechanism, which said that due to the nonequilibrium heated electronic regions from the optical penetration of the laser energy, the abrupt increase of electronic pressure exerts on the surface of the metal lattices and leads to the removal of the lattices [16–19]. Nevertheless, the mechanisms responsible for metal film thickness dependent femtosecond laser spallation and ablation are still open. The film thickness determines the available depth of thermal diffusion of the deposited laser energy [20], overall degree of heating and damping of the thermal stress induced by femtosecond laser pulse.

*Corresponding Author: zhangyu@missouri.edu

In this paper, a multiscale framework integrating the quantum mechanics (QM), molecular dynamics (MD) and two-temperature model (TTM) [21–25], is to be employed to study the femtosecond laser interaction with silver film. The spallation and ablation of silver film is observed from the atomic motion of silver atoms from MD simulation. Besides the original interatomic force in MD, an additional force acting on the nuclei is modeled by taking the electron-phonon coupled heat transfer into account.

## 2. SIMULATION METHOD

Considering the femtosecond laser radius is in the length scale of micrometer, the computational domain was treated as one dimensional in the $x$-direction. Free boundary conditions were applied on the two surfaces of the silver film, while period boundary conditions were applied on the surfaces perpendicular to $y$- and $z$-directions. The overall length of the computational domain varied with thicknesses of the five films. Five silver films with thicknesses of $392.19 nm$, $522.92\ nm$, $653.65\ nm$, $784.38\ nm$ and $915.11\ nm$ were simulated. The thickness of silver film chosen in this paper covers the cases of femtosecond laser induced the phenomena from pure spallation and the coexistence of spallation and ablation.

The volumetric energy source $S$ of the incident laser pulse at per unit time obeys the temporal Gaussian distribution [26]

$$S(x,t) = \frac{0.94 J_{abs}}{t_p L} exp(-\frac{x}{L}) exp[-2.77 \frac{(t-t_0)^2}{{t_p}^2}] \tag{1}$$

The silver film samples were initially prepared at room temperature $300\ K$ for $10\ ps$. The first $5\ ps$ canonical ensemble simulation was for the preparation of equilibrating the electron subsystem and the lattice subsystem at room temperature $300\ K$. The second $5\ ps$ microcanonical ensemble simulation was for the verification of the thermal equilibrium. The QM-MD-TTM combined simulation started at $10\ ps$. $t_p$ and $t_0$ were set as $500 fs$ and $25\ ps$, respectively. The entire simulation lasted for $200\ ps$. The energy equation of electrons is described in

$$C_e \frac{\partial T_e}{\partial t} = \nabla(k_e \nabla T_e) - G_{e-ph}(T_e - T_l) + S(x,t) \tag{2}$$

The finite difference method (FDM) is used to solve the evolution of the electron energy in Eq (2). $C_e$, $k_e$ and $G_{e-ph}$ are to be determined from QM calculation, namely,

$$\begin{cases} C_e|_{T_e} = \frac{1}{V_c} \int_{-\infty}^{\infty} \left( \frac{\partial g|_{T_e}}{\partial T_e} f|_{T_e} + g|_{T_e} \frac{\partial f|_{T_e}}{\partial T_e} \right) \varepsilon d\varepsilon \\ k_e|_{T_e} = \frac{1}{3V_c} v_F^2 \tau_e|_{T_e} \int_{-\infty}^{\infty} (\frac{\partial g|_{T_e}}{\partial T_e} f|_{T_e} + g|_{T_e} \frac{\partial f|_{T_e}}{\partial T_e}) \varepsilon d\varepsilon \\ G_{e-ph}|_{T_e} = \frac{1}{V_c} \frac{\pi \hbar k_B \lambda \langle \omega^2 \rangle|_{T_e}}{g(\varepsilon_F)|_{T_e}} \int_{-\infty}^{\infty} g|_{T_e}^2 (-\frac{\partial f|_{T_e}}{\partial \varepsilon}) d\varepsilon \end{cases} \tag{3}$$

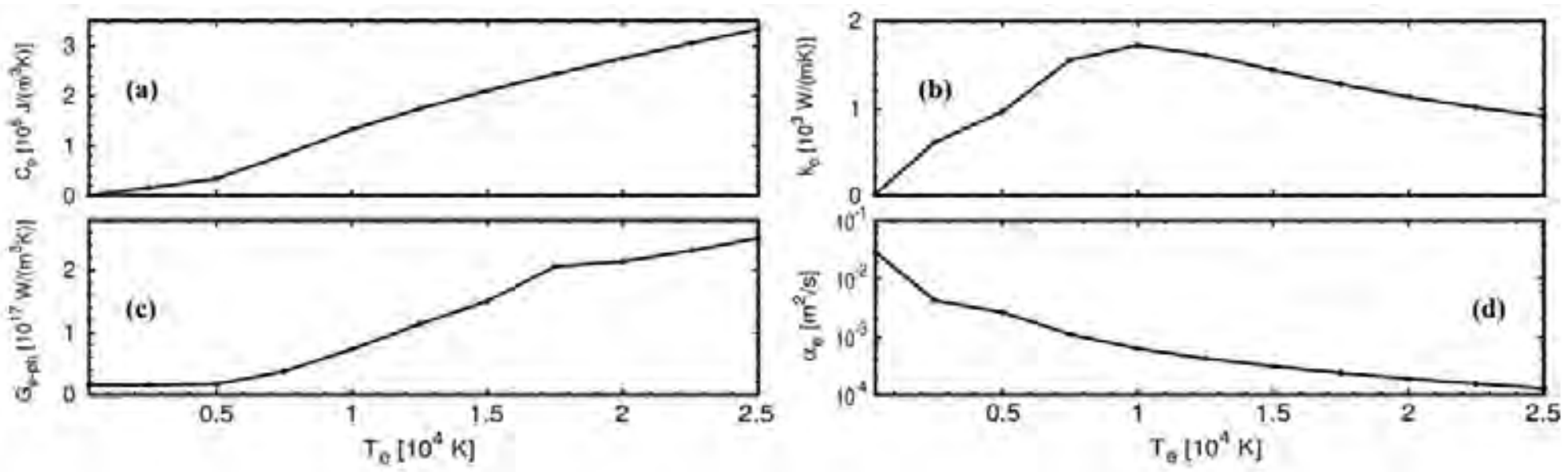

**Fig. 1** Electron temperature dependent thermophyscical variables of electron subsystem

Under femtosecond laser excitation, $g|_{T_e}$, $g(\varepsilon_F)|_{T_e}$, $f|_{T_e}$ and $\lambda\langle\omega^2\rangle|_{T_e}$ in change with different $T_e$. $\hbar$ and $k_B$ are reduced Planck constant and Boltzmann constant, respectively. Finite temperature density functional theory (FT-DFT) was implemented in calculating $g|_{T_e}$, $g(\varepsilon_F)|_{T_e}$, $f|_{T_e}$ and $\lambda\langle\omega^2\rangle|_{T_e}$. The valence electrons $4d^{10}5s^1$ was taken in FT-DFT calculation. The local density approximation (LDA) with a plane wave cutoff of 28 $eV$ was adopted in computing the exchange and correlation energy. $10 \times 10 \times 10$ Monkhorst-Pack $k$-point grids were used to sample the Brillouin zone, which had been tested to meet the convergence. The finally calculated $T_e$ dependent $C_e$, $k_e$, $G_{e-ph}$ and thermal diffusivity $\alpha_e$ (ratio of $k_e$ to $C_e$) are plotted in Fig. 1.

The electron-phonon coupled heat transfer results in an addition force acting on the nuclei. The equation of atomic motion in MD simulation is expressed in the following equation

$$m_i \frac{d^2 \boldsymbol{r}_i}{dt^2} = -\nabla U + \frac{E_{e-ph}}{\Delta t_{MD}} \frac{m_i \boldsymbol{v}_i^T}{\sum_{j=1}^{N_V} m_j \left(\boldsymbol{v}_j^T\right)^2} \tag{4}$$

According to the kinetic theory, $T_l$ is calculated from

$$T_l = \frac{1}{3} \frac{m_i v_i^2}{k_B} \tag{5}$$

In each MD time step $\Delta t_{MD}$, the thermal energy $E_{e-ph}$ transporting from the electron subsystem to the lattice subsystem is $\Delta t_{MD} \sum_{k=1}^{n_t} G_{e-ph} V_N (T_e^k - T_l) / n_t$, where $V_N$ is the volume of FDM cell. The QM-MD-TTM integrated framework is constructed by combing Eqs. (1)-(4). The simulation code is developed as an extension of the ABINIT package [27] and the TTM part in the IMD [28]. During the QM-MD-TTM integrated simulation, the time steps $\Delta t_{MD}$ and $\Delta t_{FDM}$ were set as $1 fs$ and $0.005\ fs$ to meet the von Neumann stability criterion [29] .

## 3. RESULTS AND DISCUSSION

### 3.1 Spallation Triggered for $J_{abs} = 0.1\ J/cm^2$

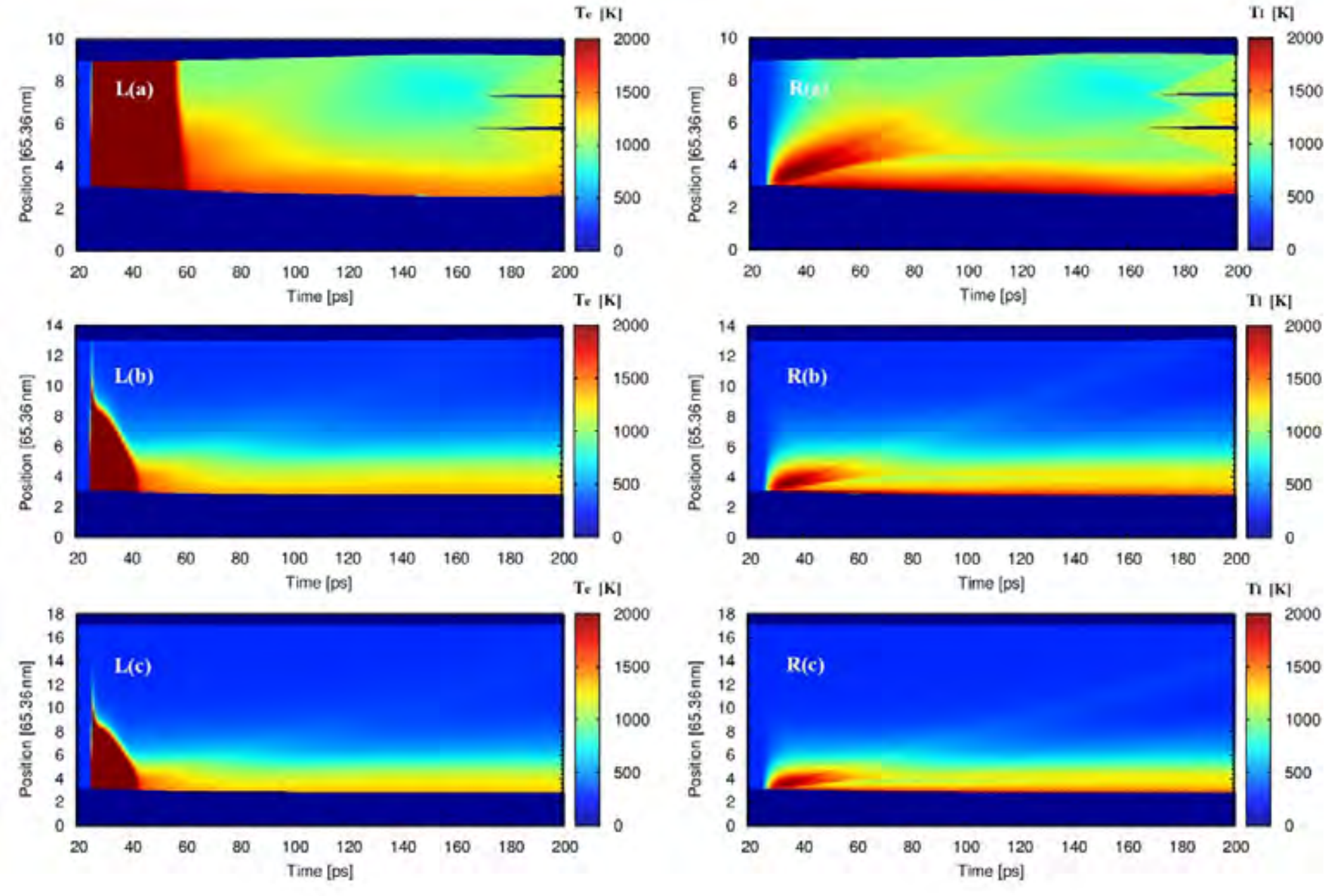

**Fig. 2** The temporal and spatial distribution of electron temperature (left) and lattice temperature (right).

When $J_{abs}$ was $0.1\ J/cm^2$, thermal melting of the front surface and laser spallation of the rear surface were seen for the film with thickness of $392.19\ nm$. Whereas, only thermal melting was found for the other four films. The temporal and spatial distribution of electron temperature $T_e$ for cases with film thicknesses of $392.19\ nm$, $653.65\ nm$ and $915.11\ nm$ are shown in Fig. 2L. The laser irradiated front surface of the silver film locates at $x = 196.09\ nm$. For the case with film thickness of $392.19\ nm$, it can been seen in Fig. 2L(a) that $T_e$ elevates to $2{,}000\ K$ (or more) throughout the silver film right after $25\ ps$, which indicates the depth of femtosecond laser heating of the electron subsystem is significantly greater than $392.19\ nm$. When the film thickness is increased to $653.65\ nm$, there is an obvious temperature difference between $T_e$ at the rear surfaces of Fig. 1(a) and Fig. 1(b) right after $25\ ps$. Furthermore, when the film thickness is increased to $915.11\ nm$, Figure 2L(c) shows that $T_e$ at the rear surface is not appreciably impacted by femtosecond incident laser heating from the front surface.

According to [30–32], thermal confinement is mathematically expressed as $t_p < \tau_{e,cond} = L^2/\alpha_e$. In this paper, by taking $\alpha_e = 6.20 \times 10^{-4}\ m^2/s$ at $T_e = 10^4\ K$ (as shown in Fig. 1(d)) and $L = 68\ nm$ in Eq. (1), $\tau_{e,cond}$ is estimated as $7.46\ ps$. Considering the laser pulse duration is $500\ fs$, which is much short than the time $\tau_{e,cond}$ cost to dissipate the absorbed laser energy via electron heat conduction. As extrapolated from Fig. 1(d), the criterion of eliminating thermal confinement corresponds $\alpha_e = 9.25 \times 10^{-3}\ m^2/s$, which requires $T_e < 2046.91\ K$. Therefore, thermal confinement exists in the shallow region below the front surface of the silver film to $\sim 457.55\ nm$. The electron thermal diffusivity $\alpha_e$ calculated in Fig. 1(d) shows that at higher $T_e$, the capability of the electron subsystem to conduct the absorbed laser energy relative its capability to reserve the laser energy becomes weaker, which further enhances the thermal confinement. Moreover, at higher $T_e$, $G_{e-ph}$ present increasing trend, which leads to the faster transfer rate of confined thermal energy from the electron subsystem to the lattice subsystem. Two splits are found at $\sim 170\ ps$ in Fig. 2L(a). Since the two splits occur at the region far behind the laser heated front surface and $T_e$ at the original points of these two points are smaller than that under the front surface, it is conjectured that the two splits are caused by mechanical reason, rather than the thermal reason.

The temporal and spatial distribution of lattice temperature $T_l$ in Fig. 2R presents the detailed information on $T_l$ evolution, which is resulted from the electron-phonon coupled heat transfer. At $t = 25\ ps$, even though the electron subsystem has been heated throughout the silver film (in Fig. 2L), the majority part of the lattice still keep cold (in Fig. 2R). Right after tens of picoseconds, the lattice subsystem is gradually heated. In Fig. 2R(a), the lattice subsystem is heated greater than $1{,}007.12\ K$ after $50\ ps$. However, owing to the sufficient thickness for the films in Figs. 2R(b) and 2R(c), the depths of heated lattice subsystem ($T_l > 1000\ K$) are limited to $150.34\ nm$ below the front surface of the silver film. It should be pointed out that there is a $T_l$ decreased region ($145\ ps < t < 165\ ps$) near the rear surface of the silver film in Fig. 2R(a). Right after the $T_l$ decreased region disappears, the rear ($x = 475.86\ nm$) split appears. Besides recording the temperature evolutions, the density spatial distribution of density is also calculated. Results show the densities of the two regions after spallation ($x = 475.86\ nm$) becomes smaller. Therefore, the decrease of both $T_e$ and $T_l$ in Fig. 2(a) are caused by transformation from kinetic energy of atoms to interatomic potential energy. However, for the other (front $x = 376.50\ nm$) split in Fig. 2R(a), $T_l$ at its starting point is not seen obvious difference from its foregoing region. Therefore, the reason leading to these two splits remains open.

**Table 1** Film thickness dependent melting depth and spallation depth.

| Case | Film Thickness ($nm$) | $L_{mel}$ ($nm$) | $T_l$ ($K$) (at $L_{mel}$) | $L_{spa}$ ($nm$) |
|---|---|---|---|---|
| 1 | 392.19 | 111.12 | 1205.65 | 183.02 |
| 2 | 522.92 | 84.97 | 1183.88 | N/A |
| 3 | 653.65 | 71.90 | 1256.67 | N/A |
| 4 | 784.38 | 75.17 | 1238.80 | N/A |
| 5 | 915.11 | 62.10 | 1239.27 | N/A |

The film thickness dependent laser melting depth $L_{mel}$ for the cases with $J_{abs} = 0.1\,J/cm^2$ are measured and listed in Table 1. The reference point of $L_{mel}$ is taken as the initial location of the front surface ($196.09\,nm$) of the laser film. The melting boundary is defined as the location where significant mass density change appears. As seen in Table 1, $L_{mel}$ decreases with the increase of film thickness. With the film thickness increases from $392.19\,nm$ to $522.92\,nm$, the greatest decrease of melt depth is seen in Table 1. When the same amount of laser energies are deposited into the five silver films with different thickness, the thicker silver film will result in smaller Kelvin degree of laser heating. $T_l$ at the melting boundary show values close to the reported melting point of silver $1234.93\,K$ [33], which verifies the validity of the current QM-MD-TTM integrated simulation. During the $200\,ps$ simulation, laser spallation is triggered only for the case with film thickness of $392.19\,nm$. From the perspective of micromachining, when $J_{abs} = 0.1\,J/cm^2$, for the silver film with thickness $392.19\,nm$ can be spalled with the steady surface left for the remained bulk film. As for the other four films, longer simulation still needs to be carried out to see whether the laser spallation will happen or not.

### 3.2 Coexistence of Spallation and Ablation Triggered for $J_{abs} = 0.3\,J/cm^2$

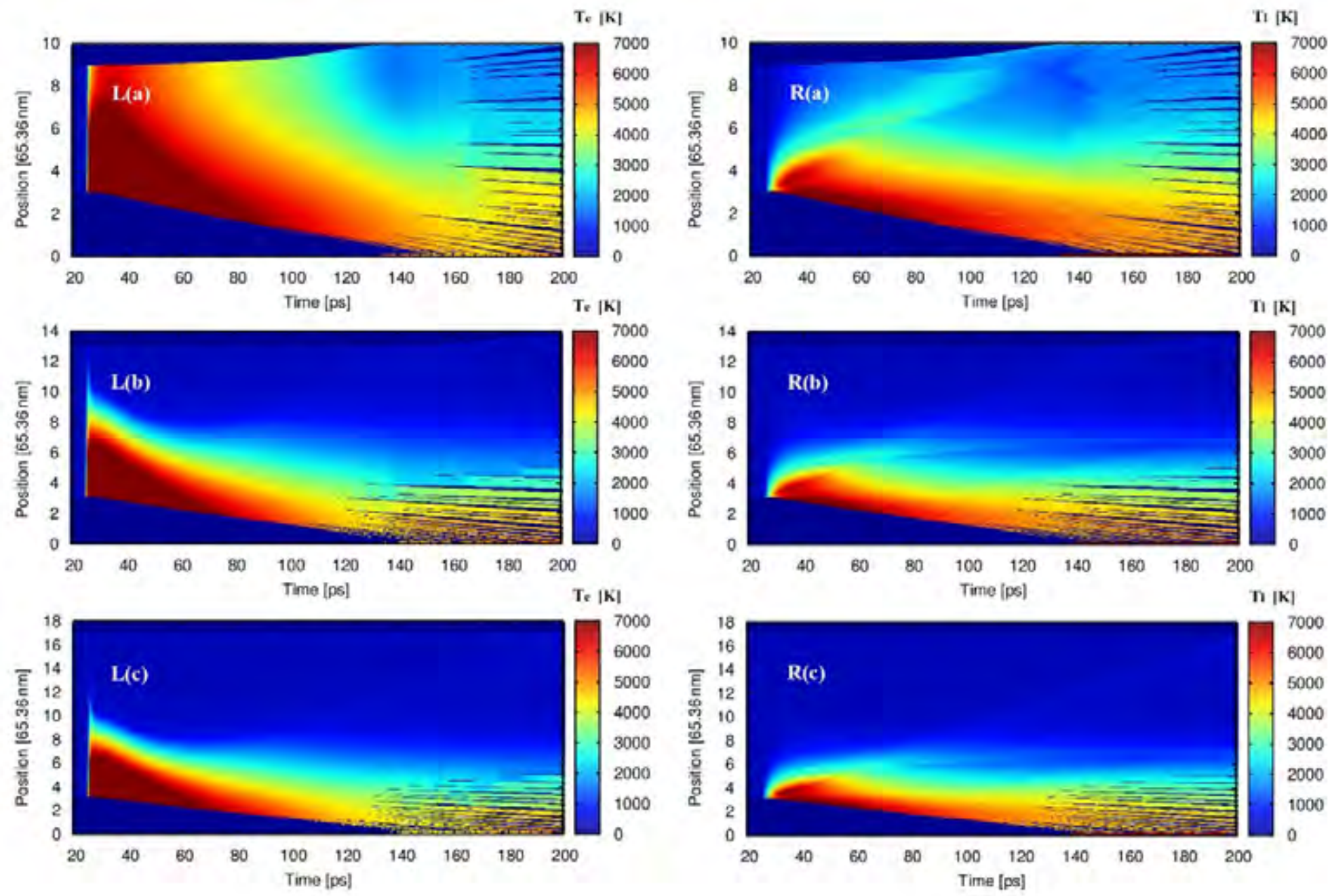

**Fig. 3** The temporal and spatial distribution of electron temperature (left) and lattice temperature (right).

When $J_{abs}$ is $0.3\,J/cm^2$, femtosecond laser induced spallation and ablation were seen for the cases with film thicknesses of $392.19\,nm$ and $522.92\,nm$. Whereas, only ablation was seen for the other three cases with film thicknesses of $653.65\,nm$, $784.38\,nm$ and $915.11\,nm$. Unlike spallation, the ablation is defined as the small segments of laser heated silver fly away from the front surface of the film.

The distributions of $T_e$ for the cases with thicknesses of $392.19\,nm$, $653.65\,nm$ and $915.11\,nm$ are plotted in Fig. 3L. Comparing with the $T_e$ distribution for the cases in Fig. 2L, it can be seen that the time cost for the high $T_e$ to get cooled down is much longer in Fig. 3L, owing to the larger amount of laser energy deposition into the silver film. Moreover, recalling $t_{e-ph}$ calculated in from ratio of $C_e/G_{e-ph}$, the greater $T_e$ leads to longer time during the process of decreasing $T_e$ by electron-phonon coupled heat transfer, which is another factor leading to the longer time cost for the cases in Fig. 3L than those in Fig. 2L. At $27\,ps$ of the cases in Fig. 3L, $T_e$ at the rear surface of the silver film are $5071.58\,K$, $354.89\,K$ and $303.01\,K$, respectively. Therefore, it can be concluded that even though the laser fluence increases, the silver film with thickness greater than $653.65\,nm$ is sufficient to keep the rear side of the film being not significantly heated. At $170\,ps$, the silver film in Fig. 3L(a) has been totally split into several segments. It should be noted that $T_e$ of these segments from the front surface to the rear surface range from $4533.94\ K$ to $1693.15\ K$. Therefore, besides the discussed mechanical reason of laser spallation at lower temperatures, the mechanism leading to those hot splits remains open.

The distributions of $T_l$ for the cases with $J_{abs} = 0.3\ J/cm^2$ are calculated and plotted in Fig. 3R. With the increase of the film thickness, the depth of heated region changes a lot. $T_l$ at $80\ ps$ at $x = 457.55\ nm$ for the three cases Figs. 3R(a)-(c) are $2817.66\ K$, $1575.14\ K$ and $1579.81\ K$, respectively, which confirms that when the film thickness is greater than $653.65\ nm$, the effect of film thickness for laser heating with $J_{abs} = 0.3\ J/cm^2$ is no longer a dominate factor. As seen in Fig. 3R(b), a $T_l$ line appears since $70\ ps$ and develops to the deeper regions with the evolution of time. With the continuous progress of simulation, $T_l$ gradually becomes in consistent with $T_e$ in Fig. 3L. The splits generated front the front surfaces of the cases in Figs. 3R(b) (corresponding to the ablation depth of $137.27\ nm$) and 3R(c) (corresponding to the ablation depth of $130.73\ nm$)show values above boiling point $2435.15\ K$ of silver [33]. Explosive boiling are found from the front surface when $25\ ps < t < 80\ ps$, which is verified by $T_l$ show values greater than $0.9$ times of the critical temperature $6{,}410.15\ K$ of silver [33]. Due to the limits to the current QM-MD-TTM simulation, the generation of plasma is not included in this paper.

**Table 2** Film thickness dependent spallation depth and ablation depth.

| Case | Film Thickness $(nm)$ | $L_{spa}\ (nm)$ | $L_{abl}\ (nm)$ |
|---|---|---|---|
| 1 | 392.19 | 451.02 | 261.46 |
| 2 | 522.92 | 535.99 | 210.80 |
| 3 | 653.65 | N/A | 137.27 |
| 4 | 784.38 | N/A | 134.00 |
| 5 | 915.11 | N/A | 130.73 |

Table 2 presents the spallation depth $L_{spa}$ and ablation depth $L_{abl}$ for the cases with $J_{abs} = 0.3\ J/cm^2$. With the increase of laser fluence from $0.1\ J/cm^2$ to $0.3\ J/cm^2$, laser spallation is also seen for the film with thickness of $522.92\ nm$ during the $200\ ps$ simulation time, which is not seen when $J_{abs} = 0.1\ J/cm^2$. The results in Table 2 indicates that when $J_{abs} = 0.3\ J/cm^2$, the laser spallation damages the silver films (with thicknesses of $392.19\ nm$ and $522.92\ nm$) into a few small segments. For the silver film without laser spallation, steady surface is seen from the front side of the remained silver film.

## 4. CONCLUSIONS

To sum up, this paper takes advantages of the highly accurate QM determination of the electron thermophysical properties, the detailed description of the laser pulse induced atomic motion and phase change process, as well as the inclusion of energy evolution of the laser energy excited electron subsystem in continuum. Laser ablation happens at much higher $T_l$ than that of laser spallation and is limited to the laser heated region under the front surface. The coexistence of ablation and spallation results in the damage of the silver film into several small segments. The size of spallation segment (pure laser spallation seen in Fig. 2(a) ) is much larger than the size of ablation segment (pure laser ablation seen in Fig. 3(c)). In addition, laser spallation can happen without accompanying melting in the ablated surface. Therefore, from the perspective of improving the quality of micromachining, to properly choose the thinner silver film enables spallation from the rear surface and keep melting at the front surface of low laser fluence heating.

## ACKNOWLEDGMENT

Support for this work by the U.S. National Science Foundation under grant number CBET- 133611, the National Natural Science Foundation of China under grant numbers 51705234, U1537202 and 2015B010132005, the Southern University of Science and Technology Presidential Postdoctoral Fellowship and the projected funded by China Postdoctoral Science Foundation under grant number 2017M612653 are gratefully acknowledged.

## NOMENCLATURE

| | | |
|---|---|---|
| $A$ | material constants describing the electron-electron scattering rate | $(s^{-1}K^{-2})$ |
| $B$ | material constants describing the electron-phonon scattering rate | $(s^{-1}K^{-1})$ |
| $f$ | Fermi-Dirac distribution function | (-) |
| $g$ | electron density of states | (-) |
| $G_{e-ph}$ | electron-phonon coupling factor | $(W/(m^3K))$ |
| $J_{abs}$ | absorbed laser fluence | $(J/cm^2)$ |
| $k_e$ | electron thermal conductivity | $(W/(mK))$ |
| $L$ | penetrating depth | $(m)$ |
| $N_v$ | number atoms in FMD cell | (-) |
| $\boldsymbol{r_i}$ | position of an atom | $(m)$ |
| $T_e^k$ | average electron temperature of computational cell per FDM time step | $(K)$ |
| $U$ | interatomic potential | $(J)$ |
| $V_c$ | volume of unit cell | $(m^3)$ |
| $V_N$ | volume of FDM cell | $(m^3)$ |
| ***Greek Letters*** | | |
| $\varepsilon$ | electron energy level | $(J)$ |
| $\mu$ | chemical potential | $(J)$ |
| $\lambda\langle\omega^2\rangle$ | second moment of the electron-phonon spectral function | $(meV^2)$ |
| $\tau_e$ | total electron scattering time | $(s)$ |
| ***Subscripts and Superscripts*** | | |
| $e$ | electron | |
| $F$ | Fermi | |
| $l$ | lattice | |
| $op$ | optical | |
| $p$ | pulse | |